\documentclass[showpacs,prb,amsmath,amssymb,twocolumn]{revtex4}

\ifx\pdftexversion\undefined
  \usepackage[dvips]{graphics}
\else
  \usepackage[pdftex]{graphics}
\fi
\usepackage{graphicx}

\begin{document}
\title{Persistent and radiation-induced currents in distorted quantum
rings}

\author{Yuriy V. Pershin}
\author{Carlo Piermarocchi}

\affiliation{\small Department of Physics and Astronomy, Michigan
State University, East Lansing, Michigan 48824-2320}

\begin{abstract} Persistent and radiation-induced currents in
distorted narrow quantum rings are theoretically investigated. We show
that ring distorsions can be described using a geometrical potential
term. We analyse the effect of this term on the current induced by a
magnetic flux (persistent current) and by a polarized coherent
electromagnetic field (radiation-induced current). The strongest
effects in persistent currents are observed for distorted rings with a
small number of electrons. The distortion smoothes the current
oscillations as a function of the magnetic flux and changes the
temperature dependence of the current amplitude. For radiation-induced
currents, the distortion induces an ac component in the current and
affects its dependence on the radiation frequency and intensity.
\end{abstract}


\maketitle

\section{Introduction} It is well known that a static magnetic flux
through a mesoscopic ring induces a dissipationless non-decaying
(persistent) current at low temperature. During the last 20 years
this persistent current has been heavily investigated both from
the theoretical and experimental
side.~\cite{ABEandPC,PCwithSO,VRWZ98,Serega,Ulloa,r1,r2,r3,r4,r5,r6,r7,r8,r9,
Mohanty99,Mohanty01,LossMartin,Szopa,magarill96,bulaev04,manolescu,per05}
In particular, theoretical investigations were focused on the
effects of electron-electron interaction, disorder,
spin-orbit,~\cite{PCwithSO} polarized nuclear
spins,~\cite{VRWZ98,Serega} and shape.~\cite{magarill96,bulaev04}
More recently, the effect of electromagnetic radiation on
mesoscopic rings has been investigated. It has been pointed out
that the persistent current can be strongly affected by
radiation~\cite{manolescu} and also that radiation can induce a
current at zero magnetic flux.~\cite{per05,matos05} In order to
break the clockwise-anticlockwise symmetry and obtain a current at
zero flux a radiation with some degree of circular polarization is
needed. This can be achieved by a superposition of pairs of
time-asymmetric, linearly cross-polarized picosecond
pulses,~\cite{matos05} or more simply by using circularly
polarized radiation.~\cite{per05} All these effects were mainly
studied in systems with perfect ring geometry, i.e. rings where
the confinement potential does not depend on the azimuthal
coordinate.  The only exceptions, to our knowledge, are
Refs.~\onlinecite{magarill96} and \onlinecite{bulaev04}, where the
persistent current in an elliptical quantum ring \cite{magarill96}
and the persistent current in a quantum ring on a surface of
constant negative curvature \cite{bulaev04} were considered.

\begin{figure}[b]
\centering \includegraphics[angle=270, width=7cm]{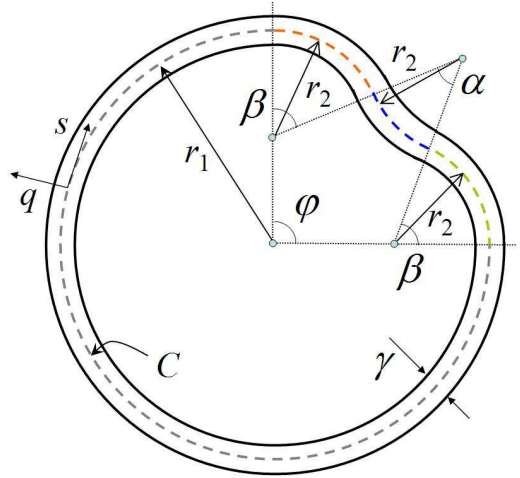}
\caption{(Color online) Distorted quantum ring consisting of a
long segment of radius $r_1$ and three short segments of radius
$r_2$.} \label{fig1}
\end{figure}

In this paper we investigate persistent and radiation-induced
currents in distorted quantum rings. We consider a narrow
distorted ring of uniform cross section lying on a plane (Fig.
\ref{fig1}). The ring can be composed by several segments of
different curvature or can have any smooth curvature profile. We
show that the curvature of the ring enters into the Scr\"odinger
equation via a geometrical potential term of the form
$V_{geom}=-\hbar^2/(8m^*R^2)$, where $R$ is the radius of
curvature. Our model differs from the one used in
Ref.~\onlinecite{magarill96}, where a quantum ring of a
non-uniform cross section was considered. By choosing a uniform
cross section the effect of the distorsion can be described in a
simpler and more transparent way.
%***********
Our paper is organized as follows. In Sect.~\ref{sec2} we derive the
Schr\"odinger equation for one electron in a distorted ring in the
presence of a magnetic flux. We show that, as in the ideal case, the
wavefunction in the distorted ring is a periodic function of the
magnetic flux $\Phi$ with period $\Phi_0$. Next, in Sect.~\ref{sec3},
we consider a model where a distorted quantum ring consists of four
constant-curvature segments. We find the energy spectrum of such a
ring and we demonstrate that the geometrical potential $V_{geom}$ in
this case opens gaps in the electron energy spectrum. Moreover, we
show that the geometrical potential can lead to bound states. The
oscillation of the persistent current, and the frequency and intensity
dependence of the radiation-induced currents in the distorted ring are
studied in Sec.~\ref{sec4}.  The results of our investigations are
summarized in Sec.~\ref{sec6}.

\section{Effective Hamiltonian} \label{sec2}
We consider one electron with
effective mass $m^*$ confined by a potential $V_\gamma$ ($\gamma$
denotes the characteristic width of $V_\gamma$) to a closed curve $C$
on a plane. A uniform magnetic field $H$ perpendicular to the plane is
applied. The Schr\"odinger equation has the form
\begin{equation}
\frac{1}{2m^*}\left( \mathbf{\hat{p}} -\frac{e}{c}\mathbf{A}
\right)^2\psi+V_\gamma\psi=E\psi \label{shr},
\end{equation}
where $ \mathbf{\hat{p}}$ is the electron momentum operator and
$\mathbf{A}(\mathbf{r})=\frac{1}{2}[\mathbf{H},\mathbf{r}]$ is the
vector potential. Using the property
$\textnormal{div}(\mathbf{A})=0$ of this gauge, Eq.~(\ref{shr})
can be rewritten as
\begin{equation}
\frac{1}{2m^*}\left( -\hbar^2 \Delta
-2\frac{e}{c}\mathbf{A}\mathbf{\hat{p}}+\frac{e^2}{c^2}\mathbf{A}^2
\right)\psi+V_\gamma\psi=E\psi \label{shr1}.
\end{equation}
Our goal is to obtain an effective one-dimensional Schr\"odinger
equation in the limit of a strong transverse confinement, i.e. in the
limit $\gamma \rightarrow 0$. We follow the approach proposed in
Ref.~\onlinecite{costa81} and subsequently used in
Ref.~\onlinecite{shevchenko01}.
%Our calculations extend previous derivations (such as reported in
%Refs. \cite{costa81,shevchenko01}) of the effective equations
%describing one-dimensional motion for the case of non-zero vector
%potential.

Let us introduce the orthonormal coordinate system ($s$, $q$),
where $s$ is the arc length parameter and $q$ is the coordinate
along the normal $\mathbf{n}(s)$. The curve $C$ is then described
by a vector $\mathbf{r}(s)$ as a function of the arc length $s$.
In a vicinity of $C$ the position is therefore is described by
\begin{equation}
\mathbf{R}(s,q)=\mathbf{r}(s)+q\mathbf{n}(s).
\end{equation}
For the sake of simplicity~\cite{shevchenko01} we assume that
$V_\gamma$ depends only on the $q$ coordinate describing the
displacement from the reference curve $C$ only.

The Laplacian $\Delta$ in the curvilinear coordinates $s$ and $q$ is
given by
\begin{equation}
\Delta_{s,q}=\frac{1}{h}\frac{\partial}{\partial
s}\frac{1}{h}\frac{\partial}{\partial
s}+\frac{1}{h}\frac{\partial}{\partial q}h
\frac{\partial}{\partial q},
\end{equation}
with
\begin{equation}
h=1-k(s)q,
\end{equation}
where $k(s)=R^{-1}(s)$ is the curvature. Using the transformation
to the new wave function $\chi(s,q)$ via
$\psi(s,q)=\chi(s,q)/\sqrt{h}$ (note, that $\chi(s,q)$ is properly
normalized), we can rewrite Eq.~(\ref{shr1}) as
\begin{widetext}
\begin{eqnarray}
\frac{1}{2m^*}\left[ -\hbar^2 \left( \frac{\partial}{\partial
s}\frac{1}{h^2}\frac{\partial}{\partial
s}-\frac{h_{ss}}{2h^3}+\frac{5h_s^2}{4h^4}+\frac{\partial^2}{\partial
q^2}+\frac{k^2}{4h^2}\right) + \right. \nonumber \\ \left.
2i\frac{e\hbar}{c}\left(A_s(s,q)\left(\frac{\partial}{\partial s }
+\frac{k_sq}{2h}\right)+A_q(s,q)\left(\frac{\partial}{\partial q
}+\frac{k}{2h}\right)\right)+\frac{e^2}{c^2}\mathbf{A}^2
\right]\chi+  V_\gamma\chi=E\chi \label{shr3}~,
\end{eqnarray}
\end{widetext}
where $h_s=\partial h/\partial s$, $h_{ss}=\partial^2 h/\partial s^2$,
$k_s=\partial k/\partial s$, $A_q(s,q)$ and $A_s(s,q)$ are the
components of $\mathbf{A}$ along the $q$ and $s$ directions. Next, we
make the substitution $\chi=\textnormal{exp}\left[i\frac{e}{\hbar
c}\int\limits_0^qA_q(s,q')dq'\right]\tilde{\chi}$, and expand $h$,
$A_q(s,q)$, $A_s(s,q)$ in series in $q$ keeping only the zero-order
terms in $q$, as in Refs.~\onlinecite{costa81,shevchenko01}. The
Schr\"odinger equation~(\ref{shr3}) can then be easily separated by
setting $\tilde{\chi}(s,q)=\nu(q)\phi(s)$. The usual procedure yields
\begin{equation}
-\frac{\hbar^2}{2m^*}\frac{\partial^2 \nu}{\partial q^2
}+V_\gamma\nu=E_t\nu, \label{shrt}
\end{equation}
\begin{eqnarray}
\left[-\frac{\hbar^2}{2m^*} \frac{\partial^2}{\partial s^2}+
i\frac{e\hbar}{m^*c}A_s(s,0)\frac{\partial}{\partial s }
-\frac{\hbar^2k^2}{8m^*}-\frac{ie\hbar}{2m^*c} \times \right.
\nonumber
\\ \left. \frac{\partial
A_q (s,0) }{\partial q }+i\frac{e\hbar}{m^*c}A_q(s,0)\frac{k}{2}
+\frac{e^2}{2m^*c^2}A_s^2(s,0)\right]\phi=E_l\phi . \,
\label{shr5}
\end{eqnarray}
In order to further simplify Eq.~(\ref{shr5}), we perform the
transformation  $\phi(s)=\textnormal{exp}\left[i\frac{e}{\hbar
c}\int\limits_0^sA_s(s',0)ds'\right]\tilde{\phi}(s)$, which gives
\begin{eqnarray}
\left[-\frac{\hbar^2}{2m^*}\frac{\partial^2 }{\partial s^2}
-\frac{\hbar^2k^2}{8m^*}-\frac{ie\hbar}{2m^*c}\frac{\partial
A_s(s,0)}{\partial s }- \right. \nonumber
\\ \left.
\frac{ie\hbar}{2m^*c}a_1(s)+i\frac{e\hbar}{m^*c}a_0\frac{k}{2}
\right]\tilde{\phi}(s)=E_t\tilde{\phi}(s). \label{shr9}
\end{eqnarray}
Notice that in the curvilinear coordinates ($s$, $q$) the divergence
of $\mathbf{A}$ is given by
\begin{eqnarray}
\textnormal{div}\mathbf{A}=
\frac{1}{h}\frac{\partial}{\partial_s}\left(\frac{1}{h}A_s\right)+\frac{1}{h}\frac{\partial}{\partial
q }\left(hA_q\right)\approx \nonumber \\
\frac{\partial}{\partial_s}A_s+\frac{\partial}{\partial q }A_q-kA_q=0.
\end{eqnarray}
Consequently, Eq. (\ref{shr9}) reduces to
\begin{equation}
-\frac{\hbar^2}{2m^*}\frac{\partial^2 \tilde{\phi}}{\partial s^2}
-\frac{\hbar^2k^2(s)}{8m^*}\tilde{\phi}=E_l\tilde{\phi}.
\label{shrl}
\end{equation}
Therefore, we have derived two decoupled equations: one describing the
transverse confinement of electrons in the ring (Eq.~(\ref{shrt})),
and the second describing the longitudinal motion of the electron in
the ring (Eq.~(\ref{shrl})).  The vector potential $\mathbf{A}$ does
not explicitly appear in these two equations.  However it will appear
in the solution of Eq.~(\ref{shrl}) because of the boundary conditions
on $\tilde{\phi}$ specified below. The spectrum of Eq.~(\ref{shrt})
depends on the particular shape of the confinement potential
$V_\gamma$. In this paper we assume that the electrons occupy only the
lowest subband of the transversal confinement. Therefore, the position
of this energy subband is not important. The curvature of C enters into
Eq. (\ref{shrl}) through the geometrical potential term $-\hbar^2
k(s)^2/8 m^*$.

The boundary conditions for $\tilde{\phi}$ are obtained from the
requirements of continuity of the wave function $\phi(s)$ and its
derivative, i.e. $\phi(0)=\phi(L)$, $\partial \phi(0)/\partial
s=\partial \phi(L)/\partial s$ ($L$ is the ring circumference).
Using Stokes' theorem we finally obtain
\begin{equation}
\tilde{\phi}(0)=e^{i2\pi\frac{\Phi}{\Phi_0}}\tilde{\phi}(L), \quad
\frac{\partial \tilde{\phi}(0)}{\partial s}=\frac{\partial
\tilde{\phi} (L)}{\partial s}. \label{boundcond}
\end{equation}
Here $\Phi$ is the magnetic flux through the area confined by C
and $\Phi_0$ is the magnetic flux quantum. Eqs.~(\ref{boundcond})
imply that all equilibrium physical properties of a narrow closed
loop are periodic in $\Phi$ with period $\Phi_0$, as in the case
of a perfect ring.~\cite{r3}

\section{Electron energy spectrum} \label{sec3}
In this section we consider the electron spectrum in a ring with a
dent-type distortion as shown in Fig.~\ref{fig1}. Geometrically, the
curve C of such a ring consists of four smoothly connected
circular segments. The radius of the long segment is $r_1$, the
three short segments have the same radius $r_2$ (we assume here that
$r_2<r_1$).
%It can be shown that such a curve can be simply
%geometrically constructed.
The angles $\alpha$ and $\beta$ are related to $\varphi$ (for the
definition of these angles see Fig.  \ref{fig1}) as
$\alpha=2\textnormal{Arcsin}\left[\sin(\varphi
/2)(r_1-r_2)/(2r_2)\right]$, $\beta=\varphi /2+\alpha /2$.  The
advantage of a distortion with constant curvature segments is a
constant geometrical potential in each segment. The picture is even
more simple since the geometrical potential does not depend on the
direction of bending, i.e.  $V_{geom}$ is the same for the three short
segments in Fig.~\ref{fig1}. Therefore, we write Eq.~(\ref{shrl}) for
the long segment and three short segments as
\begin{eqnarray}
-\frac{\hbar^2}{2m^*}\frac{\partial^2 \phi_1}{\partial
s^2}-U_0\phi_1=E_l\phi_1 \qquad \textnormal{for $0<s<l$,}
\label{shred1} \\ -\frac{\hbar^2}{2m^*}\frac{\partial^2
\phi_2}{\partial s^2}=E_l\phi_2 \qquad \textnormal{for $l<s<L$}
\label{shred2},
\end{eqnarray}
where $l=\left( 2\beta+\alpha\right)r_2$ is the total length of
the short segments, $L=(2\pi-\varphi)r_1+l$,
$U_0=\frac{\hbar^2}{8m^*}(1/r_2^2-1/r_1^2)$.  The general solution
of Eqs.~(\ref{shred1}, \ref{shred2}) reads
\begin{eqnarray}
\phi_1=b_1e^{ik_1s}+b_2e^{-ik_1s}, \\
\phi_2=c_1e^{ik_2s}+c_2e^{-ik_2s},
\end{eqnarray}
where $k_1=\sqrt{\frac{2m^*}{\hbar}(E_l+U_0)}$ and
$k_2=\sqrt{\frac{2m^*}{\hbar}E_l}$. The wave functions $\phi_{1(2)}$
are connected at $s=l$ via $\phi_1(l)=\phi_2(l)$, $\partial
\phi_1(l)/\partial s=\partial \phi_2(l)/\partial s$ and at $l=0,L$
via Eqs. (\ref{boundcond}). From these boundary conditions we
obtain a transcendental equation defining the energy spectrum for
unbound states $E_l>0$
\begin{eqnarray}
2\cos(2\pi\frac{\Phi}{\Phi_0})+
\left[\frac{k_1}{k_2}+\frac{k_2}{k_1}\right]
\sin(k_1l)\sin(k_2(L-l))- \nonumber \\ 2\cos(k_1l)\cos(k_2(L-l))=0
\label{unbstates}
\end{eqnarray}
and for bound states $-U_0<E_l<0$
\begin{eqnarray}
2\cos\left(2\pi\frac{\Phi}{\Phi_0}\right)+
\left[\frac{k_1}{\tilde{k}_2}-\frac{\tilde{k}_2}{k_1}\right]
\sin(k_1l)\sinh(\tilde{k}_2(L-l))- \nonumber
\\ 2\cos(k_1l)\cosh(\tilde{k}_2(L-l))=0. \,\,\, \label{bstates}
\end{eqnarray}
Here, $\tilde{k}_2=\sqrt{\frac{2m^*}{\hbar}(-E_l)}$.
\begin{figure}
\centering
\includegraphics[angle=270, width=8cm]{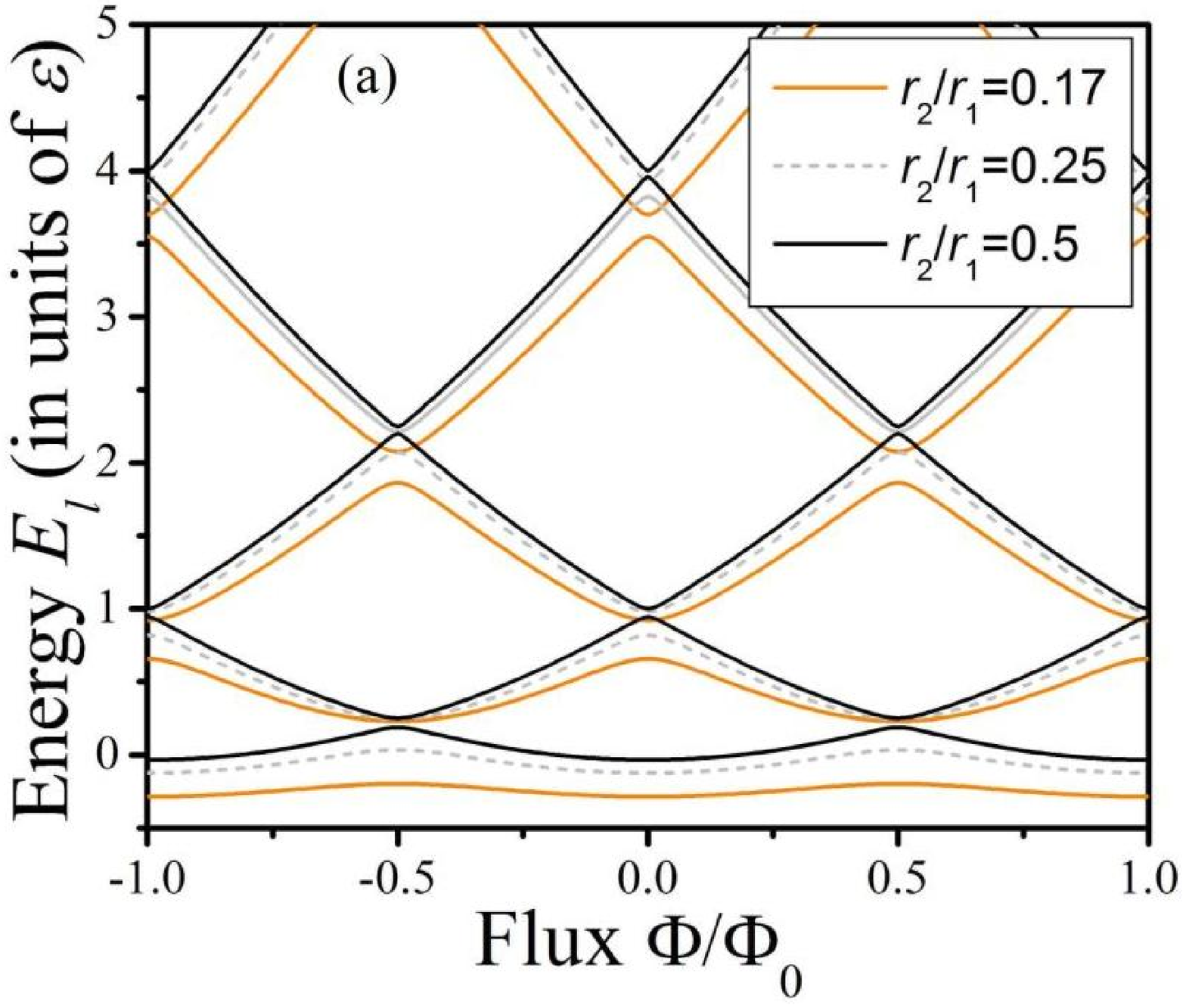}
\includegraphics[angle=-90,width=8cm]{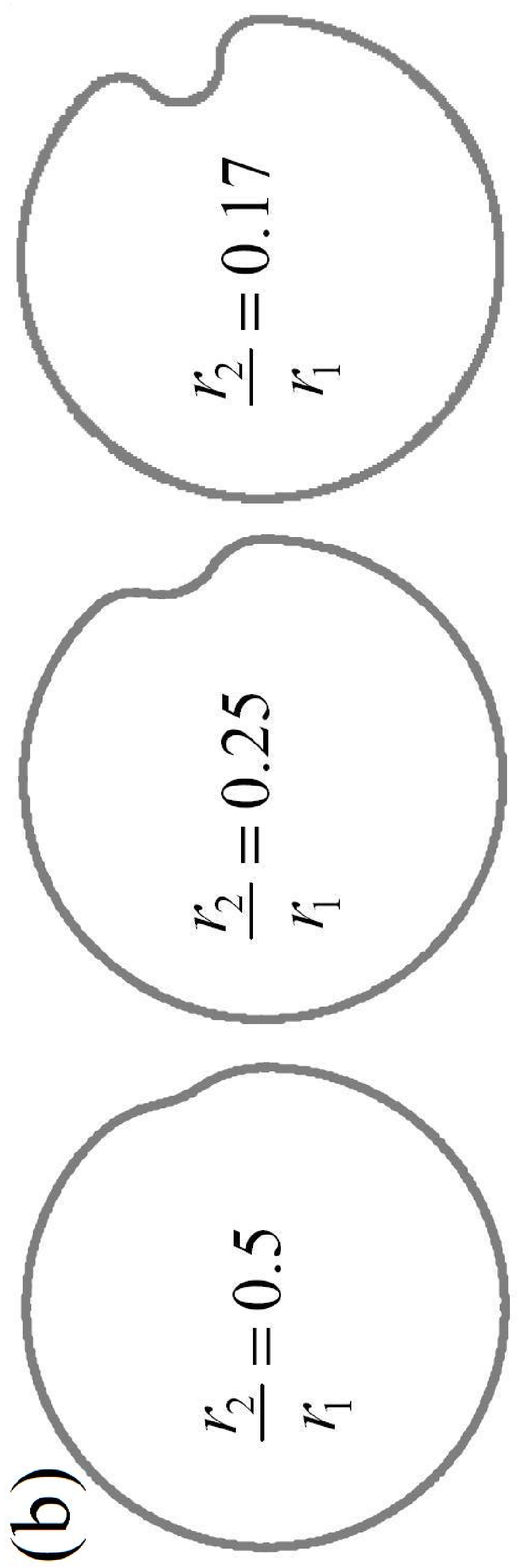}
\caption{(Color online) (a) Electron energy levels (in units of
$\varepsilon=\hbar^2/(2m^*r_1^2)$) as a function of flux in
quantum rings with different degrees of distortion at $\varphi=\pi
/4$. (b) Geometrical shape of the distorted quantum rings whose
electron energy levels are shown in (a).} \label{fig2}
\end{figure}

The calculated energy levels for weakly and strongly distorted rings
as a function of the flux are given in Fig.~\ref{fig2}. It is well
known that in perfect rings the energy levels are intersecting
parabolas. In distorted rings, gaps are opened at the points of
intersection of the parabolas. This effect is qualitatively similar to
the effect of disorder.~\cite{r3} We emphasize that for a fixed radius
of distortion, the gap decreases for larger values of the intersection
point energy. At a fixed point of intersection, the gap increases by
decreasing the radius of distortion (i.e. the gap is larger in more
distorted rings). Notice finally that due to the distortion the
effective circumference of the ring increases. This produces negative
shifts of the energy levels which are larger at higher energy.
%(It can be seen, for example, following
%the decrease of the intersection points energy in more distorted
%rings). This effect is related to increase of the ring
%circumference $L$ with increase of distortion at a fixed $\phi$.
The qualitatively new feature of the spectrum is the presence of bound
states with $E_l<0$. Similar bound states were already discussed in
elliptical quantum rings.~\cite{magarill96} It is interesting that the
transition from unbound to bound states in the ring is smooth: the
shallow bound states are still sensitive to the magnetic flux
(Fig. \ref{fig2}, $r_2/r_1=0.25$). In contrast, deep bound states have
a weak sensitivity to the magnetic flux (Fig.~\ref{fig2},
$r_2/r_1=0.17$). Correspondingly, there is a finite contribution to
the persistent current from the shallow bound states, while the
contribution from the deep bound states is small.

\section{ Effect of the distortion on the current} \label{sec4}
\subsection{Persistent currents induced by a magnetic flux}
At non-zero temperature $T$, the current in the ring is given by
\begin{equation}
I=-\frac{\partial F}{\partial \Phi}.  \label{eqforcurrent}
\end{equation}
Here, the free energy $F=-k_BT\sum\limits_{n} ln\left( 1+\exp{\frac{%
\mu-E_n}{k_BT}}\right)$, $k_B$ is the Boltzman constant, and $T$ is
the temperature. We consider a system with a fixed number of spinless
electrons $N$. The chemical potential $\mu$ that enters into $F$ is
determined by the equation
\begin{equation}
\sum_{n}\frac{1}{1+e^{\frac{E_{n}-\mu }{k_{B}T}}}=N.
\label{eqformu}
\end{equation}

\begin{figure}
\centering
\includegraphics[angle=270, width=8cm]{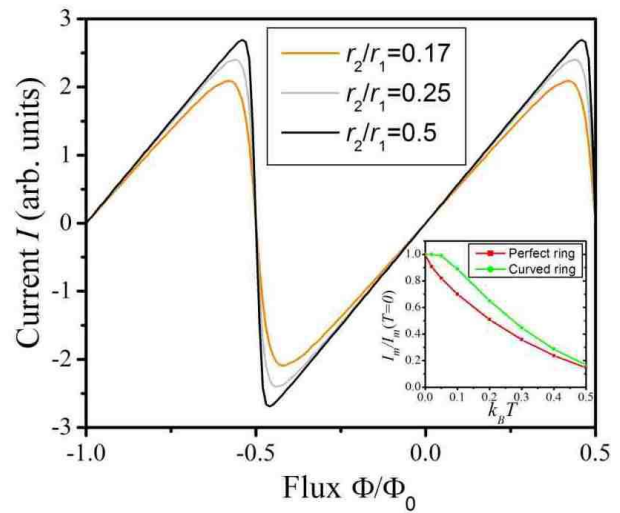}
\caption{(Color online) Persistent current in the ring at $T=0$,
$\varphi=\pi /4$, $N=3$. Inset: temperature dependence of the
persistent current amplitude in the perfect and distorted
($\varphi=\pi /4$, $r_2/r_1=0.17$) rings. $k_BT$ is in units of
$\varepsilon$.} \label{fig3}
\end{figure}

In Fig.~(\ref{fig3}), we show the effect of distortion on the
persistent current in quantum rings with three electrons. These
results were obtained numerically using Eqs.~(\ref{eqformu}) and
(\ref{eqforcurrent}) with the energy spectrum determined from
Eqs.~(\ref{unbstates}) and (\ref{bstates}). At zero temperature the
persistent current oscillations in a perfect ring have a saw-tooth
form. The distortion of the ring produces a smoothing of the
oscillations due to the opening of energy gaps at the intersection
points. Fig.~(\ref{fig3}) shows that the smoothing increases for
larger distortions. Notice that the persistent current as a function
of the magnetic field flux in the distorted ring at $T=0$ looks
similar to the persistent current in a perfect ring at $T>0$.

The distortion of the ring changes the temperature dependence of the
current amplitude. The temperature dependence of the persistent
current and its amplitude in a perfect ring with a fixed number of
electrons at low temperatures was derived in~\cite{PV} using a
two-level approach.~\cite{Vagner83} It was found that at low
temperatures the persistent current can be written as~\cite{PV}
\begin{equation}
I=\frac{2N\varepsilon }{\Phi_0}\left[ \frac {\sinh \gamma}{1+\cosh
\gamma}-2 \left( \frac{\Phi}{\Phi_0}- \frac{\delta}{2}\right) \right],
\label{PC}
\end{equation}
where $\gamma=N \varepsilon (\Phi/ \Phi_0-\delta/2)/(k_B T) $,
$\varepsilon=\hbar^2/(2m^*r^2)$,  $\delta =0$ if $N$ is even and
$\delta =1$ if $N$ is odd, and the overall factor two takes into
account the spin degeneracy. The temperature dependence of the
amplitude of the persistent current oscillations is given by
\begin{eqnarray}
I_{max}=\frac{2N\varepsilon }{\Phi_0}\left( \sqrt{1-4\frac{k_BT}{N\varepsilon%
}} - \right.  \nonumber \\ \left. 2 \frac{k_BT}{N\varepsilon}
\textnormal{arccosh} \left[ \frac{M \varepsilon }{2 k_BT}-1\right]
\right).  \label{PCmax}
\end{eqnarray}
The comparison of the temperature dependence of the persistent current
amplitude in the perfect and distorted rings is shown in the inset of
Fig.~\ref{fig3}. While the persistent current amplitude in the perfect
ring starts to decrease at $T=0$, the temperature dependence in the
distorted ring shows an activation energy behavior in the vicinity of
$T=0$ due to the gaps in the energy spectrum.

In quantum rings with many electrons the main contribution to the
persistent current is due to the electrons near the Fermi level.
Since the energy gap disappears for large values of the intersection
point energy, the persistent current at $T=0$ will not be
smoothed. Consequently, the persistent current in a distorted ring
with a large number of electrons will be as in a perfect ring with a
radius $r^*=L/(2\pi)$ in a weaker (for the case of a dent)
magnetic field $B^*=BS/S^*$, where $S$ is the area of the distorted
ring and $S^*=2\pi r^{*2}$.

\subsection{Current induced by circularly polarized radiation} \label{sec5}
In a recent paper,~\cite{per05} a novel mechanism for current
generation in quantum rings was proposed. It was suggested that in
the presence of a circularly-polarized continuous-wave (cw)
radiation the light-dressed ground state of the ring is
characterized by a non-zero current. The purpose of this Section
is to study the influence of the ring distortion on this
radiation-induced current.

Let us consider an electron confined in a distorted ring in the
presence of circularly-polarized cw radiation. The single electron
Hamiltonian in the dipole approximation reads
\begin{equation}
H=H_0+V(t)=-\frac{\hbar^2}{2m^*} \frac{\partial ^2}{\partial
s^2}+U_{geom}(s)+e\mathbf{E}(t)\mathbf{r}(s),
\end{equation}
where $\mathbf{E}(t)=E_0\cos(\omega t)\hat{x}\pm E_0\cos(\omega
t)\hat{x}$ is the circularly-polarized electric field, $E_0$ is its
amplitude, and $\pm$ corresponds to $\sigma _\pm$ radiation.  The
distorted quantum ring is considered again as made of four constant
curvature segments, which allows us to use the energy spectrum and
wave functions of $H_0$ obtained in Sec.~\ref{sec3} at
$\Phi=0$. Assuming that the radiation frequency is close to the
transition between the ground and two first excited levels, we
restrict our attention only to these three levels, with energies given
by $E_0$, $E_1$ and $E_2$.

The external radiation causes transitions between these levels.
The electron dynamics in the ring can be conveniently described
using the a density matrix approach similar to the one used in
Ref.~\onlinecite{takagahara} for quantum dots. The evolution of
density matrix $\rho$ is given by
\begin{equation}
i\hbar \dot{\rho}=[H, \rho]- \Gamma \{\rho\},
\end{equation}
where $\Gamma \{\rho\}$ represents a relaxation terms. In
the rotating wave approximation the corresponding equations for the density matrix elements are
\begin{widetext}
\begin{eqnarray}
\dot{\rho}_{00}=v_{01-}\tilde{\rho}_{10}+v_{02-}\tilde{\rho}_{20}-\tilde{\rho}_{01}v_{10+}-
\tilde{\rho}_{02}v_{20+}+\kappa_{20}\rho_{22}+\kappa_{10}\rho_{11},\label{r00}
\\
\dot{\rho}_{11}=v_{10+}\tilde{\rho}_{01}-\tilde{\rho}_{10}v_{01-}-\kappa_{10}\rho_{11}+\kappa_{21}\rho_{22},
\\
\dot{\rho}_{22}=v_{20+}\tilde{\rho}_{02}-\tilde{\rho}_{20}v_{02-}-\kappa_{20}\rho_{22}-\kappa_{21}\rho_{22},
\\
\dot{\tilde{\rho}}_{01}=\frac{ E_0-E_1 +\hbar \omega}{i\hbar}
\tilde{\rho}_{01}+v_{01-}\rho_{11}+v_{02-}\rho_{21}-\rho_{00}v_{01-}-\gamma_{01}
\tilde{\rho}_{01,} \label{r01}
\\
\dot{\tilde{\rho}}_{02}=\frac{ E_0-E_2 +\hbar \omega}{i\hbar}
\tilde{\rho}_{02}+v_{01-}\rho_{12}+v_{02-}\rho_{22}-\rho_{00}v_{02-}-\gamma_{02}
\tilde{\rho}_{02}, \label{r02}
\\
\dot{\rho}_{12}=\frac{ E_1-E_2}{i\hbar}
\rho_{12}+v_{10+}\tilde{\rho}_{02}-\tilde{\rho}_{10}v_{02-}-
\gamma_{12} \rho_{12}, \label{r12}
\\
\dot{\tilde{\rho}}_{10}=\frac{ E_1-E_0 -\hbar \omega}{i\hbar}
\tilde{\rho}_{10}+v_{10+}\rho_{00}-\rho_{11}v_{10+}-\rho_{12}v_{20+}-\gamma_{10}
\tilde{\rho}_{10}, \label{r10}
\\
\dot{\tilde{\rho}}_{20}=\frac{ E_2-E_0 -\hbar \omega}{i\hbar}
\tilde{\rho}_{20}+v_{20+}\rho_{00}-\rho_{21}v_{10+}-\rho_{22}v_{20+}-\gamma_{20}
\tilde{\rho}_{20}, \label{r20}
\\
\dot{\rho}_{21}=\frac{ E_2-E_1}{i\hbar}
\rho_{21}+v_{20+}\tilde{\rho}_{01}-\tilde{\rho}_{20}v_{01-}-
\gamma_{21} \rho_{21}. \label{r21}
\end{eqnarray}
\end{widetext}
Here, the transformations $\rho_{01}=e^{i\omega
t}\tilde{\rho}_{01}$, $\rho_{02}=e^{i\omega t}\tilde{\rho}_{02}$,
$\rho_{10}=e^{-i\omega t}\tilde{\rho}_{10}$,
$\rho_{20}=e^{-i\omega t}\tilde{\rho}_{20}$ were used,
$\kappa_{ij}$ is the relaxation rate of diagonal density matrix
elements, $\gamma_{i,j}$ is the dephasing rate of the off-diagonal
coherences $\rho_{ij}$. $v_{ij \pm }= \overline{ \langle i | V(t)
| j \rangle e^{\pm i\omega t}/(i \hbar)}$, where $\overline{ (...)
}$ denotes an averaging over a period of $V(t)$. For example, in
the case of $\sigma_+$ radiation we obtain
\begin{figure}[tb]
\centering
\includegraphics[angle=-90,width=8cm]{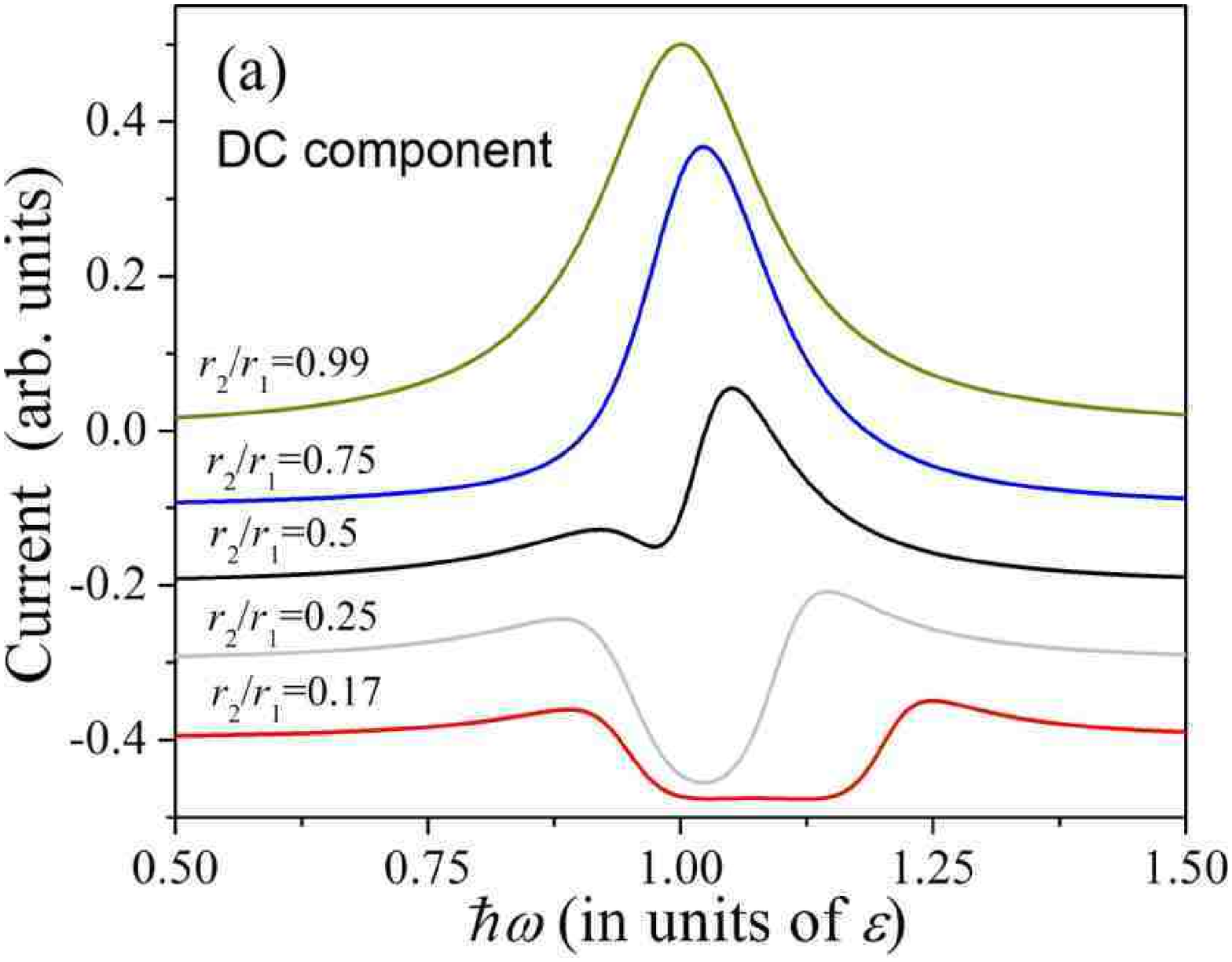}
\includegraphics[angle=-90,width=8cm]{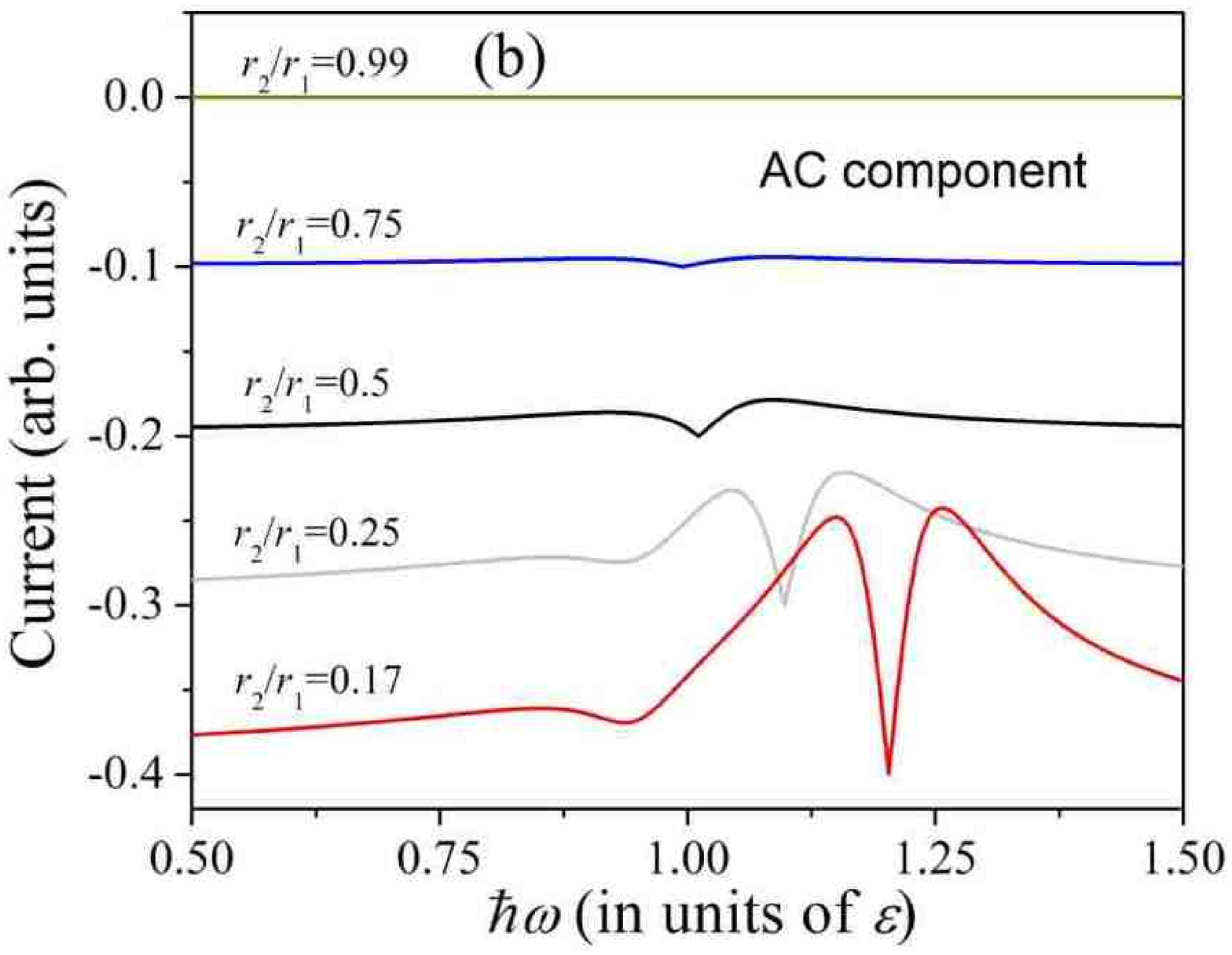}
\caption{(Color online) (a) DC component and (b) the amplitude of
the AC component of the radiation-induced current vs photon energy
for different distorted rings with the distortion parameter
$\varphi=\pi /4$ at high radiation power. Curves other than
$r_2/r_1=0.99$ were displaced for clarity. These plots were
obtained using the parameters values $\kappa_{i,j}=0.001 \hbar
/\varepsilon$, $\gamma_{i,j}=0.001 \hbar /\varepsilon$,
$eE_0r_1/\varepsilon=0.1$.} \label{fig4}
\end{figure}
\begin{figure}[tb]
\centering
\includegraphics[angle=-90,width=8cm]{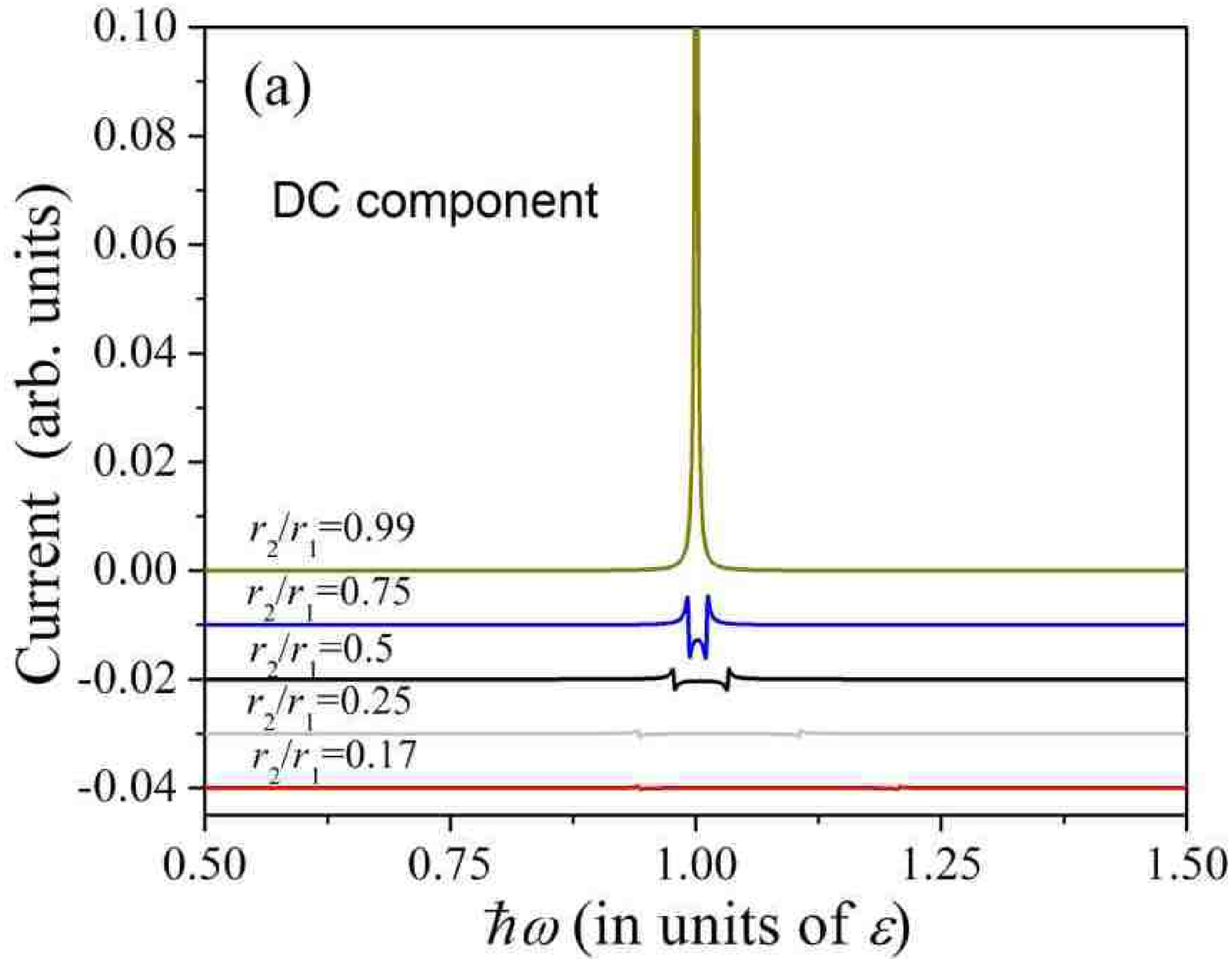}
\includegraphics[angle=-90,width=8cm]{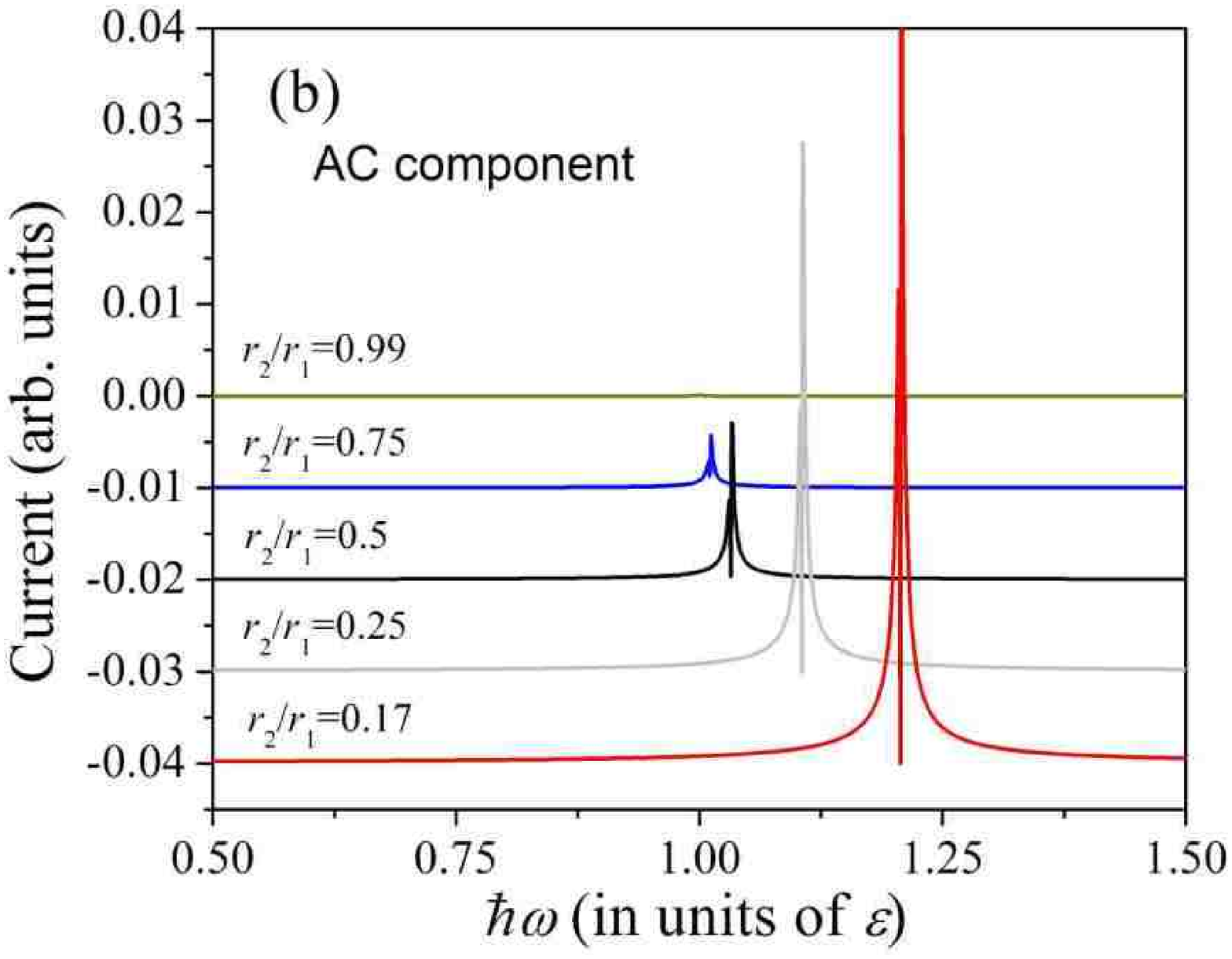}
\caption{(Color online) (a) DC component and (b) the amplitude of
the AC component of the radiation-induced current vs photon energy
at low radiation power. All calculation parameters are the same
with those in Fig. \ref{fig4} except $eE_0r_1/\varepsilon=0.001$.}
\label{fig5}
\end{figure}
\begin{eqnarray}
v_{01-}= \frac{1}{i \hbar}\frac{e E_0}{2} \langle 0|  x
+\frac{y}{i}   | 1\rangle,
\\
v_{10+}= \frac{1}{i \hbar}  \frac{e E_0}{2} \langle 1| x
-\frac{y}{i}   | 0\rangle.
\end{eqnarray}

The persistent current is calculated using $I=\textnormal{Tr}\left[
\rho \hat{j}\right]$, where $\hat{j}$ is the standard quantum
mechanical current operator, and $\rho$ is the steady-state solution
of Eqs. (\ref{r00})-(\ref{r21}). We find that in the distorted rings
the current operator matrix has a form
\begin{equation}
\hat{j}=\left( \begin{array}{ccc} 0 & 0 & j_{02} \\ 0 & 0 & j_{12}
\\ j_{20} & j_{12} & 0 \end{array} \right),
\end{equation}
with $j_{ij}=j_{ji}^*$. Correspondingly, the persistent current is
given by
\begin{equation}
I=2\textnormal{Re}\left( \rho_{21}j_{12}+\rho_{20}j_{02}
\right)=2\textnormal{Re}\left( \rho_{21}j_{12}+e^{-i\omega t
}\tilde{\rho}_{20}j_{02} \right). \label{pcd}
\end{equation}
The first term in the right hand side of Eq.~(\ref{pcd}) is
time-independent and will be referred to as the DC component of the
current, the second term in the right hand side of Eq.~(\ref{pcd}) is
the AC component of the current. In a perfect ring $j_{20}=0$, thus an
AC component in the current is a signature of ring distortion.

Figs.~\ref{fig4} and \ref{fig5} show the DC component
$2\textnormal{Re}\left( \rho_{21}j_{12} \right)$ and the amplitude of
the AC component $2|\textnormal{Re}\left( \tilde{\rho}_{20}j_{02}
\right)|$ of the radiation induced current for two values of the
radiation intensity and different distortion radii. The exact
steady state solutions of Eqs.~(\ref{r00})-(\ref{r21}) for these plots
were found numerically. In the case of a high excitation power (Fig.
\ref{fig4}) broad current peaks are observed. In the almost perfect
ring ($r_2/r_1=0.99$ curve in Fig.~\ref{fig4}(a)) the DC component has
a single resonance peak. As the distortion degree increases, this peak
shifts to a higher energy and its amplitude decreases (see
$r_2/r_1=0.75$ curve). For a stronger distortion ($r_2/r_1=0.5$) a
second peak appears at a lower energy, which is related to the lower
energy splitted level $E_1$. At $r_2/r_1=0.25$ and $r_2/r_1=0.17$ an
additional negative current peak is observed. The amplitude of the AC
component of the persistent current is zero in the perfect ring. This
amplitude becomes different than zero in distorted quantum rings with
a maximum located in the region of the $E_2-E_0$ resonance
(Fig. \ref{fig4}).  The complex dependence of the DC and AC persistent
current components on the radiation frequency indicates that
significant quantum-interference effects are occurring.

The current peaks are narrower in the case of low radiation power
(see Fig.~\ref{fig5}). Fig.~\ref{fig5}(a) shows that the DC
component is suppressed in quantum rings with strong distortion.  In
contrast, the amplitude of the AC component becomes non-zero in the
distorted rings and increases with the distortion (Fig.
\ref{fig5}(b)). The maximum of the AC component is located in the
region of the $E_2-E_0$ resonance, as in the case of the high
radiation power.  We note that the sign change of
$\textnormal{Re}\left( \tilde{\rho}_{20}j_{02} \right)$ is responsible
for the vertical lines in the peak centers in Fig.~\ref{fig5}(b).

In the regime of low radiation power we can find an approximate
solution of Eqs.~(\ref{r00})-(\ref{r21}). Eq.~(\ref{r20}) gives
the following expression for $\tilde{\rho}_{20}$ in the first
order in $E_0$:
\begin{equation}
\tilde{\rho}_{20}=-\frac{i\hbar}{ E_2-E_0 -\hbar
\omega-i\hbar\gamma_{20}} v_{20+}\rho_{00}. \label{r20appr}
\end{equation}
Similarly, from Eq. (\ref{r01}) $\tilde {\rho}_{01}$ in the first
order in $E_0$ can be found. This expression for
$\tilde {\rho}_{01}$ together with Eq.~(\ref{r20appr}) and Eq.~(\ref{r21})
yields in the second order in $E_0$
\begin{eqnarray}
\rho_{21}=\frac{i \hbar}{E_2-E_1-i\hbar\gamma_{21}}
\left[\frac{i\hbar}{ E_2-E_0 -\hbar \omega-i\hbar\gamma_{20}}+
\right. \nonumber \\ \left. \frac{i\hbar}{ E_1-E_0 -\hbar
\omega-i\hbar\gamma_{01}} \right] v_{01-} v_{20+}\rho_{00}~.
\label{r21appr}
\end{eqnarray}
 We have found that the persistent current components calculated from
Eqs. (\ref{pcd}), (\ref{r20appr}), (\ref{r21appr}) with
$\rho_{0,0}=1$ perfectly coincides with the persistent current
components calculated numerically in Fig. ~\ref{fig5}. In the case
of a perfect ring the current can be seen as a ${\mathcal X}_2$
effect. The presence of a distorsion induces a ${\mathcal X}_1$
term which corresponds to the AC component.

\section{Conclusions} \label{sec6}
In conclusion, we have investigated persistent and
radiation-induced currents in quantum rings with distortions. We
have derived an effective Schr\"odinger equation describing
electrons in a narrow distorted quantum ring (closed loop) in the
presence of an external magnetic field flux. We have shown that
the electron energy spectrum is a periodic function of the
magnetic flux. The ring curvature enters into the effective
equations through a geometrical potential term. We have solved the
equations in the case of a distorted ring consisting of four
constant-curvature segments.  We have considered the effect of the
ring distortion on the magnetic flux-induced and radiation-induced
currents. It was found that the effect on the flux-induced
persistent current is more pronounced in quantum rings with a
small number of electrons and lower chemical potential. The gaps
at the points of intersection of the energy levels lead to a
smoothing of the persistent current oscillations and to a
different temperature dependence. The persistent current in a
distorted ring with a large number of electrons behaves like in a
perfect ring with a different radius in a renormalized magnetic
field.

We have also found that the ring distortion affects radiation-induced
currents. Using a density matrix approach and the rotating wave
approximation, it was found that the current in distorted quantum
rings acquires an AC component, in addition to the DC component
characteristic of perfect rings. The frequency dependence of the DC
component is modified by the distortion and shows several peaks of
different sign.  The frequency of the AC component is equal to the
radiation frequency while its amplitude increases with the
distortion. Finally, we remark that the non-trivial dependence of the
DC persistent current component on the radiation frequency can be
useful for quantum control schemes involving localized spins, as
suggested in Ref.~\onlinecite{per05}.

We thank Prof. M. Dykman for many fruitful discussions.  This research
was supported by the National Science Foundation, Grant NSF
DMR-0312491.

\end{document}